\documentclass[12pt]{article}
\usepackage{graphicx}
\begin{document}
\setcounter{section}{0}
\setcounter{equation}{0}
\noindent 
\begin{center}
{ \Large The effect of vacuum polarisation on muon-proton 
scattering at small energies and angles}
\end{center}

\vspace{1.0cm}
\noindent 
\begin{center}
A. Gashi$^a$, G. Rasche$^a$$^\dagger$, W.S. Woolcock$^b$
\end{center}

\vspace{0.3cm}
\noindent
\begin{center}

  \noindent\textit{{$^{a}$Institut f\"{u}r Theoretische Physik der
  Universit\"{a}t,\\
  	Winterthurerstrasse 190, CH-8057 Z\"{u}rich, Switzerland \\
  	$^{b}$ Department of Theoretical Physics, IAS, \\
  	The Australian National University, Canberra, ACT 0200, Australia}}
  \\
\end{center}

\noindent 
\vspace{0.7cm}

We give a compact expression for the unpolarised differential 
cross section for muon-proton scattering in the one photon exchange approximation. 
The effect of adding the vacuum polarisation amplitude to the no-spin-flip amplitude for 
one photon exchange is calculated at small energies and scattering angles and is found to be 
negligible for present experiments.

\vspace{0.7cm}

\vspace{0.7cm}

\vspace{0.7cm}
\noindent $^\dagger$ Corresponding author. Electronic mail: rasche@physik.unizh.ch;
  Tel: +41 1 635 5810;  Fax: +41 1 635 5704

\newpage

A group at TRIUMF is currently running an experiment to measure the differential cross sections for 
$\pi^+p$ and $\pi^-p$ elastic scattering at small energies(pion laboratory kinetic energy in the range 20 MeV 
to 60 MeV) and at small angles (from $2^{\circ}$ to $30^{\circ}$) [1]. To fix the absolute values of the desired cross sections accurately, 
simultaneous measurements are being made of the cross sections for muon-proton scattering. This cross section is given 
to a very good approximation by calculating it using the one photon exchange ($1\gamma E$) amplitude. Two photon 
exchange diagrams can certainly be ignored, but at the small energies and angles at which the experiments are being done 
it is necessary to check whether the inclusion of vacuum polarisation makes a change to the results obtained using 
the $1\gamma E$ approximation that needs to be taken into account in the present experiment. This note reports 
the results of our calculations of this effect.

It turns out that the correct formula for the unpolarised differential cross section (average over initial spins, sum over final spins) 
for muon-proton scattering in the $1\gamma E$ approximation is not readily accessible in the literature; an unnecessarily complicated form is 
in private circulation. The full calculation for the laboratory frame has been given in Ref.[2], together with results 
for the polarisation parameters and references to earlier work. The unpolarised cross section is the quantity $I_0$ in Eq.(4.6) of Ref.[2]. 
In the following we will put it in a neater form which involves the electromagnetic form factors $F_1$, $F_2$ of the proton. 
This form can be readily programmed and used to determine the average muon-proton cross section across the target in a typical experiment, 
where the muons lose a non-negligible amount of energy in the target and an average cross section is what is actually measured. 

A muon, with mass $\mu$ and incident laboratory kinetic energy $T$, is scattered by a proton (mass $m$) at rest, through an angle $\theta$ 
in the laboratory. The total initial energy and laboratory momentum of the incident muon are given by 
\begin{equation}
\omega=\mu+T \, , \, l^2=T(2\mu+T)
\end{equation}
The Mandelstam variables $s$, $t$ are given by
\begin{equation}
s=(m+\mu)^2+2mT,
\end{equation}
\begin{equation}
t=\frac{-2ml^2\{\omega \sin^2 \theta +m-\cos \theta \sqrt {m^2 -\mu^2 \sin^2 \theta}\} }{s+l^2 \sin^2 \theta}   \,.
\end{equation}
The muon emerges with a laboratory momentum $l'$, where 
\begin{equation}
\frac{l'}{l}=\frac{(\omega+m) \sqrt {m^2-\mu^2\sin^2\theta}+(m\omega +\mu^2)\cos \theta}{s+l^2\sin^2 \theta}\,.
\end{equation}

The unpolarised differential cross section in the centre-of-momentum (c.m.) frame is given in its most economical form in 
terms of the Dirac and Pauli form factors $F_1$ and $F_2$ of the proton, normalised so that 
\[
F_1(0)=F_2(0)=1\,.
\]
The anomalous magnetic moment of the proton then appears explicitly and is taken from Ref.[3]. The result is 
\begin{eqnarray}
\frac{d\sigma}{d\Omega_{cm}}=\frac{1}{2s} \frac{\alpha^2}{t^2} \{ (F_1(t))^2 [2(s-m^2-\mu^2)^2+2st+t^2] \nonumber \\
+2\kappa F_1(t)F_2(t)t(t+2\mu^2) \nonumber \\
+\frac{t\kappa^2(F_2(t))^2}{2m^2} [t(2m^2+\mu^2-s)-\lambda(s,m^2,\mu^2)] \} \, ,
\end{eqnarray}
where
\begin{equation}
\lambda(s,m^2,\mu^2)=s^2-2s(m^2+\mu^2)+(m^2-\mu^2)^2
\end{equation}
and $s$, $t$ are defined in Eqs.(2) and (3). The conversion to the differential 
cross section in the laboratory frame is via a kinematical factor:
\begin{equation}
\frac{d\sigma}{d\Omega_{lab}}=(\frac{l'}{l})^2\frac{s}{m\sqrt{m^2-\mu^2\sin^2\theta}}\frac{d\sigma}{d\Omega_{cm}}\, ,
\end{equation}
where $\frac{l'}{l}$ is given by Eq.(4).

The other aim of this paper, beside giving the results in Eqs.(5) and (7), is to report our calculations of the effect of adding the 
vacuum polarisation amplitude to the no-spin-flip $1\gamma E$ amplitude. This amplitude in the c.m frame we denote by $f$; 
$f$ is real and contributes $f^2$ to $\frac{d\sigma}{d\Omega_{cm}}$. The rest of $\frac{d\sigma}{d\Omega_{cm}}$ comes from the 
5 spin-flip amplitudes, whose form is of no interest to us here. It is completely sufficient to use the lowest order approximation to 
the vacuum polarisation amplitude $f_{vp}$ in the c.m. frame; $f_{vp}$ is a no-spin-flip amplitude and is real. Moreover, 
the effect of the proton form factor on $f_{vp}$ is negligible. The effect of inclusion of $f_{vp}$ is to increase 
$\frac{d\sigma}{d\Omega_{cm}}$ by 
\[
(f+f_{vp})^2 -f^2=f_{vp}(2f+f_{vp})\, ;
\]
$f$ and $f_{vp}$ have the same sign. The quantity of interest to us is the proportional increase $r_{vp}$: 
\begin{equation}
r_{vp}=\frac{f_{vp}(2f+f_{vp})}{\frac{d\sigma}{d\Omega_{cm}}}\,.
\end{equation}
Since the conversion to $\frac{d\sigma}{d\Omega_{lab}}$ is via the purely kinematical factor of Eq.(7), $r_{vp}$ 
in Eq.(8) is also the proportional increase in $\frac{d\sigma}{d\Omega_{lab}}$ due to vacuum polarisation. 

The expression for $f$ comes from a rather tedious calculation starting from Eq.(2.17) of Ref.[2], but with the matrix elements of the 
currents evaluated in the c.m. frame. The expression for $f_{vp}$ can be taken from Ref.[4] or Ref.[5]. For $\mu^+p$ scattering $f$ 
and $f_{vp}$ are, in the present notation,
\begin{eqnarray}
f=\frac{\alpha}{Wt}\{F_1(t)[s-m^2-\mu^2+t(\frac{W}{W+m+\mu}+\frac{1}{2}\frac{s+(m-\mu)^2}{s-(m-\mu)^2}) \nonumber \\
+\frac{st^2}{2(W+m+\mu)^2(s-(m-\mu)^2)}] \nonumber \\
+\frac{\kappa tF_2(t)}{2m}[W-m+\frac{st}{(W+m+\mu)(s-(m-\mu)^2)}] \}\, ,
\end{eqnarray}
\begin{equation}
f_{vp}=\frac{\alpha^2(s-m^2-\mu^2)}{3\pi Wt}F(t)\, ,
\end{equation}
where $W=\sqrt{s}$ and
\begin{equation}
F(t)=-\frac{5}{3}+X+(1+X)^{\frac{1}{2}}(1-\frac{1}{2}X)\ln \{\frac{(1+X)^{\frac{1}{2}}+1}{(1+X)^{\frac{1}{2}}-1}\}\, ,
\end{equation}
\begin{equation}
X=-\frac{4m_e^2}{t}\, ,
\end{equation}
$m_e$ being the electron mass. For $\mu^-p$ scattering both $f$ and $f_{vp}$ change sign, so that $r_{vp}$ remains 
unchanged.

For our calculations we used the form factors 
\begin{equation}
F_1(t)=F_2(t)=(1-t/\Lambda^2)^{-2}\, ,
\end{equation}
with $\Lambda=805$ MeV, which corresponds to the measured charge radius of the proton [6]. The values of $r_{vp}$ 
were calculated for $\theta\leq 30^{\circ}$ at values of $T$ between 20 MeV and 60 MeV. The value of $r_{vp}$ for a given angle $\theta$ increases very
slowly as $T$ increases. We give the results at  20 MeV, 40 MeV and 60 MeV in Fig.1. Contrary to expectations, they show that $r_{vp}$ increases with the angle as
well as with 
energy, though of course the unpolarised cross section itself falls with energy for a fixed angle $\theta$ and 
falls very rapidly as one moves away from the forward direction. 

The important conclusion is that $r_{vp}$ is in fact much smaller than the best statistical error that could 
realistically be achieved in the counting rate for the relevant scattering experiments ($\mu^{\pm}p$ and 
$\pi^{\pm}p$). The effect of vacuum polarisation is a correction similar in size to what one would expect from 
two photon exchange, around $1\%$ or less. Both effects can be safely ignored when using the $\mu^{\pm}p$ differential 
cross section to fix the absolute values of the differential cross sections for 
$\pi^{\pm}p$ elastic scattering at small energies and angles.

{\bf Acknowledgements.} 
We thank the group performing the TRIUMF experiment 778 for suggesting the problem and the 
Swiss National Foundation for financial support.

\begin{figure}
  \begin{center}
    \includegraphics[height=0.55\textheight,angle=0]
      {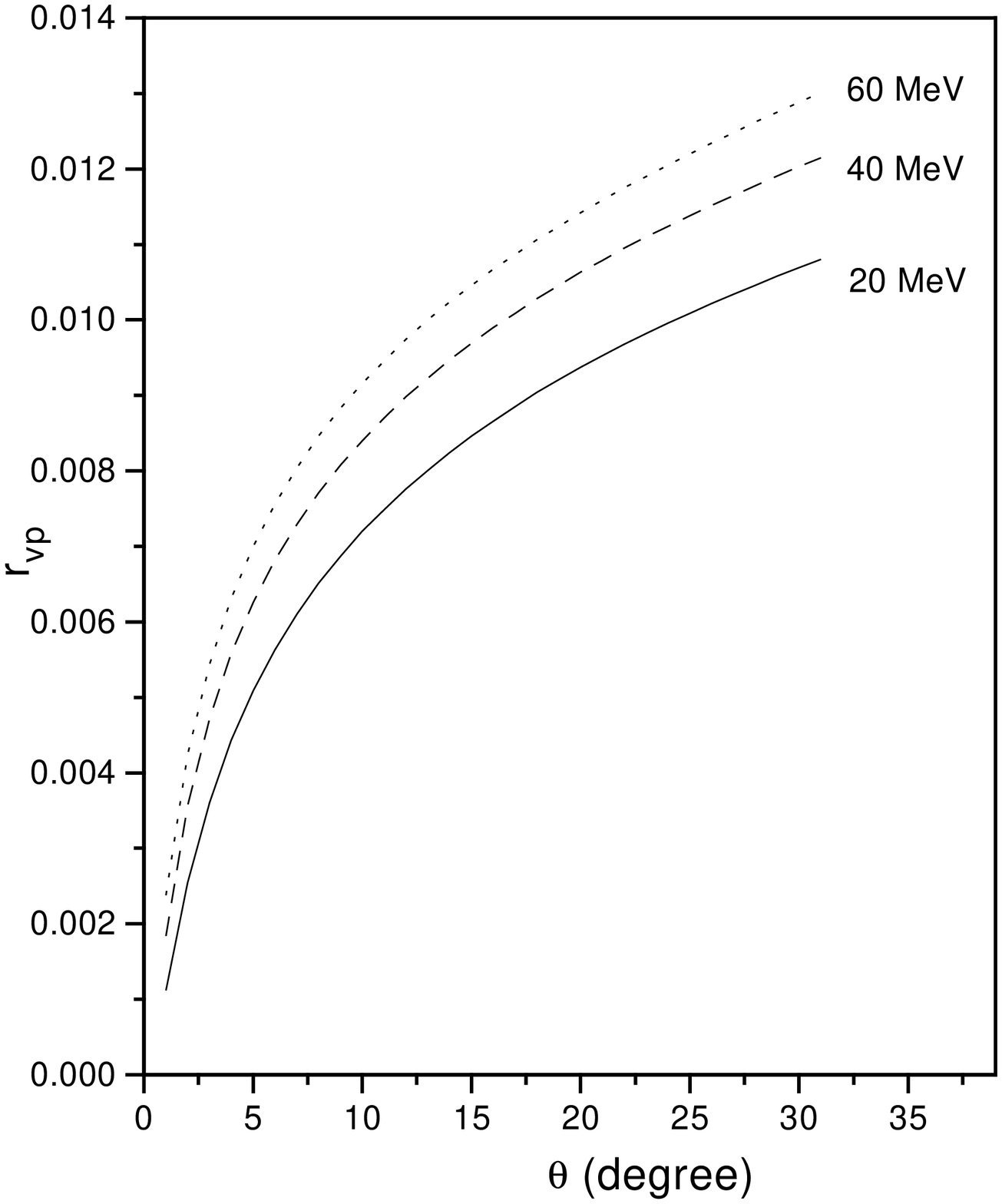}
      \caption{The proportional increase $r_{vp}$ in the differential cross
      section for $\mu^{\pm}p$ scattering due to vacuum polarisation at lab angles $\theta\leq 30^{\circ}$ and muon lab kinetic energies 20 MeV, 40 MeV and 60 MeV. }
    \end{center}
\end{figure}

\end{document}